\documentclass[conference,10pt,a4paper,oneside,twocolumn,final]{IEEEtran}
\IEEEoverridecommandlockouts

\usepackage{cite}
\usepackage{multirow}
\usepackage{booktabs}
\usepackage{comment}
\usepackage{amsmath,amssymb,amsfonts}
\usepackage{algorithmic}
\usepackage{graphicx}
\usepackage{textcomp}
\usepackage{xcolor}
\usepackage{hyperref}
\usepackage{lipsum}
\ifCLASSOPTIONcompsoc
    \usepackage[caption=false, font=normalsize, labelfont=sf, textfont=sf]{subfig}
\else
\usepackage[caption=false, font=footnotesize]{subfig}
\fi
\def\BibTeX{{\rm B\kern-.05em{\sc i\kern-.025em b}\kern-.08em
    T\kern-.1667em\lower.7ex\hbox{E}\kern-.125emX}}

\begin{document}

\title{An Efficient Quality Metric for Video Frame Interpolation Based on Motion-Field Divergence\\

\thanks{This publication has emanated from research jointly funded by Taighde Éireann – Research Ireland, and Overcast HQ under Grant number EPSPG/2023/1515. Thanks also to Neil Birkbeck and Balu Adsumilli from the YouTube Video Algorithms team for their helpful comments and supporting this work.}
}

\author{

    \IEEEauthorblockN{Conall Daly, Darren Ramsook, and Anil Kokaram}
    \IEEEauthorblockA{\textit{Sigmedia Group, School of Engineering} \\
    \textit{Trinity College Dublin}\\
            Ireland \\
            \{dalyc21, ramsookd, anil.kokaram\}@tcd.ie
    }
}

\maketitle

\begin{abstract}

Video frame interpolation is a fundamental tool for temporal video enhancement, but existing quality metrics struggle to evaluate the perceptual impact of interpolation artefacts effectively. Metrics like PSNR, SSIM and LPIPS ignore temporal coherence. State-of-the-art quality metrics tailored towards video frame interpolation, like FloLPIPS, have been developed but suffer from computational inefficiency that limits their practical application. We present PSNR\textsubscript{DIV}, a novel full-reference quality metric that enhances PSNR through motion divergence weighting, a technique adapted from archival film restoration where it was developed to detect temporal inconsistencies. Our approach highlights singularities in motion fields which is then used to weight image errors.  Evaluation on the BVI-VFI dataset (180 sequences across multiple frame rates, resolutions and interpolation methods) shows PSNR\textsubscript{DIV} achieves statistically significant improvements: +0.09 Pearson Linear Correlation Coefficient over FloLPIPS, while being 2.5$\times$ faster and using 4$\times$ less memory. Performance remains consistent across all content categories and are robust to the motion estimator used. The efficiency and accuracy of PSNR\textsubscript{DIV} enables fast quality evaluation and practical use as a loss function for training neural networks for video frame interpolation tasks. An implementation of our metric is available at \url{www.github.com/conalld/psnr-div}.

\end{abstract}


\section{Introduction}
Video frame interpolation (VFI) is a widely studied problem in video processing, aimed at generating intermediate frames between existing ones in a sequence. By increasing the effective frame rate, VFI enhances visual quality and enables applications such as slow-motion effects \cite{paliwal2020slowmotion}, efficient rendering of animated content \cite{briedis2021neural}, and broadcast frame-rate conversion \cite{guo2017frame}. In recent years, significant advancements have been made in VFI techniques, driven by sophisticated motion estimation algorithms \cite{teed2020raft} and deep neural networks (DNN). DNN based VFI algorithms are capable of accurately predicting intricate, non-linear motion patterns. They also handle occlusions more effectively than classical methods. Techniques used for DNN-based VFI have included motion-based\cite{liu2017video}, kernel-based \cite{niklaus2017video,danier2022stmfnet}, and transformer networks \cite{shi2022video}. Despite these significant advancements in interpolation quality, evaluating and comparing VFI methods remains challenging. 

Traditional metrics such as PSNR and SSIM are commonly applied to video quality assessment by averaging frame-level scores. However, this approach fails to account for temporal information, leading to poor correlation with perceptual judgments in video frame interpolation (VFI). While perceptual metrics like LPIPS \cite{zhang2018unreasonable} improve upon traditional methods by leveraging deep learning to compare structural and textural differences, they still operate on a per-frame basis. As a result, they overlook motion consistency which is a critical aspect of perceived video quality \cite{mackin2017high}. This shortcoming is particularly evident in VFI, where temporal artifacts significantly impact subjective evaluations \cite{danier2022subjective}, as well as in diffusion-based video generation, where maintaining frame coherence remains challenging \cite{ge2024content}.

To address this limitation, recent adaptations have incorporated temporal awareness into quality assessment. For instance, FloLPIPS \cite{danier2022flolpips} enhances LPIPS by weighting motion errors to highlight motion-related distortions. While this improves detection of temporal artifacts, its high computational cost (reported as 5.5$\times$ slower than standard LPIPS) makes it impractical for training DNNs.

To address these limitations we propose a novel motion-focused quality metric for VFI evaluation that is simpler and correlates more closely to subjective scores than FloLPIPS. Our key contributions are as follows.

\begin{enumerate}

    \item We propose a metric which takes a novel approach to weighting image errors based on blotch detection work in the field of archival footage restoration.
    
    \item This metric only requires the motion field for one of the compared video sequences resulting in a reduction in computational load.
    
    \item We demonstrate that our metric can outperform FloLPIPS with better correlation to subjective scores.
    
\end{enumerate}

The rest of the paper is structured as follows: Section \ref{sec:background} reviews background of quality metrics for VFI. Section \ref{sec:prop_metric} details our proposed VFI quality metric. Section \ref{setup} details the BVI-VFI subjective dataset we use to evaluate our metric and the other metrics we compare against. Section \ref{sec:eval} discusses the results of our evaluation and looks at the robustness of our metric to motion fields generated by different estimation algorithms.
  
\section{Background}
\label{sec:background}
This section reviews previous works on quality measurement for VFI and how temporal consistency measures can be leveraged as a video quality measure for VFI.

\subsection{Video Frame Interpolation Quality Metrics}
\label{sec:vfi-qual-metrics}
Video quality metrics are categorized as no-reference (NR), reduced-reference (RR), or full-reference (FR). Our work focuses on FR metrics, which compare interpolated sequences against a ground-truth reference. Metric performance is typically evaluated by correlation with human subjective scores, using statistical measures such as Pearson’s Linear Correlation Coefficient (PLCC), Spearman’s Rank Correlation Coefficient (SRCC), Kendall's Rank Correlation Coefficient (KRCC) and Root Mean Squared Error (RMSE), as standardized by VQEG \cite{vqeg2000final}.

Recently, video quality metrics increasingly leverage spatio-temporal features \cite{han2025full,ccokmez2025clip}. ST-GREED \cite{madhusudana2021stgreed} analyses spatial/temporal band-pass coefficients via Generalized Gaussian Distributions (GGD), measuring entropy differences between reference and interpolated videos to detect distortions from frame rate changes and compression. In contrast, Wu \textit{et al.} \cite{wu2019quality} directly use estimated motion, tracking keypoint trajectories and extracting motion field patches around them. By comparing histograms of these patches between reference and interpolated sequences, they quantify temporal consistency deviations.

However, general-purpose video metrics often correlate poorly with subjective scores for VFI-specific artefacts \cite{men2020visual}, prompting the development of bespoke solutions. Hou \textit{et al.} \cite{hou2022perceptual} employ a Swin Transformer-based architecture trained directly on subjective scores from their own custom dataset (VFIPS). It achieves good correlation in terms of PLCC (0.87) and SRCC (0.79). However, it is computationally intensive, requiring 2.5 GB of GPU memory to process a 256×256 12-frame sequence. 

Danier \textit{et al.} propose FloLPIPS, a video quality metric that extends LPIPS by incorporating temporal distortion sensitivity. LPIPS computes perceptual differences between reference ($\mathbf{I}$) and interpolated ($\hat{\mathbf{I}}$) frames using deep feature maps $\phi_l(\hat{\mathbf{I}})$, $\phi_l(\mathbf{I})$ from layer $l$ of a pre-trained convolutional neural network. For each spatial position $\mathbf{x}$, LPIPS calculates $\|w_l \odot (\phi_l(\hat{\mathbf{I}})-\phi_l(\mathbf{I}))\|_2^2$ across channels, where $w_l$ is a weighting to emphasise or de-emphasise the features of layer $l$. It then averages these differences spatially. FloLPIPS enhances LPIPS by introducing motion error weighting. For consecutive frames $\hat{\mathbf{I}}_{n-1}$, $\hat{\mathbf{I}}_n$ and $\mathbf{I}_{n-1}$, $\mathbf{I}_n$, it computes motion fields $\hat{\mathbf{F}}$ and $\mathbf{F}$ using a pre-trained estimator. The end-point-error (EPE) between these motion fields is calculated.

\begin{equation}
\Delta \mathbf{F}(\mathbf{x}) = \|\hat{\mathbf{F}}(\mathbf{x}) - \mathbf{F}(\mathbf{x})\|_2
\end{equation} 

This EPE map is normalised spatially to produce distortion-sensitive weights.

\begin{equation}
\label{eq:flow-weight}
\mathbf{w}(\mathbf{x}) = \frac{\Delta \mathbf{F}(\mathbf{x})}{\sum_{\mathbf{x}} \Delta \mathbf{F}(\mathbf{x})}
\end{equation}

This weights LPIPS's uniform spatial averaging, emphasizing regions with motion inconsistencies. The final FloLPIPS score aggregates the weighted feature differences across layers and frames.

\begin{equation}
\text{FloLPIPS} = \sum_l \sum_{\mathbf{x}} \mathbf{w}(\mathbf{x}) \|w_l \odot (\phi_l(\hat{\mathbf{I}})-\phi_l(\mathbf{I}))\|_2^2
\end{equation}

FloLPIPS achieves scores of 0.71 in terms of PLCC and 0.68 in terms of SRCC on the BVI-VFI dataset. As mentioned earlier, FloLPIPS is quite computationally intensive. FloLPIPS runtime is reported as 332 ms for a 1080p frame whereas LPIPS is 59 ms, making FloLPIPS 5.5$\times$ slower. FloLPIPS power comes from using EPE as a measure of temporal consistency. As such, we propose to use a simpler more computationally efficient measure of temporal consistency.

\subsection{Divergence as a Measure of Temporal Consistency}
Temporal consistency is critical for perceived video quality \cite{pahalawatta2009motion}, distortions causing motion discontinuities are of particular interest for VFI. Our metric adapts techniques from archival film restoration, where pathological motion (PM) \cite{kokaram1998motion} characterised by irregular, non-smooth motion like fast deformations or intermittent movements pose a key challenge. PM introduces intensity discontinuities resembling physical film artefacts (scratches/blotches), leading to false alarms for traditional displaced frame difference (DFD) detectors \cite{rares2001statistical}. These methods often misclassify rapid motion or complex textures as artefacts, because their motion field patterns mimic true damage.

Corrigan \textit{et al.} \cite{corrigan2006pathological} addressed this by introducing motion field divergence to spatially identify PM. Divergence quantifies local motion irregularities: high absolute values indicate abrupt vector changes, violating smooth motion assumptions. By integrating divergence over a five-frame window, their method distinguishes PM from true artefacts, even when temporal profiles overlap, reducing false alarms while preserving detection in stable regions. This spatial approach complements DFD methods, robustly flagging PM.

We repurpose divergence as a computationally efficient metric for temporal consistency. Unlike EPE, which requires two motion fields for comparison, divergence can highlight motion inconsistencies within a single motion field. This approach is particularly suited to VFI, where interpolation artefacts often manifest as non-smooth motion fields.

\section{Our Proposed Metric}
\label{sec:prop_metric}

\begin{figure*}[]
    \centering
    \includegraphics[width=0.7\linewidth]{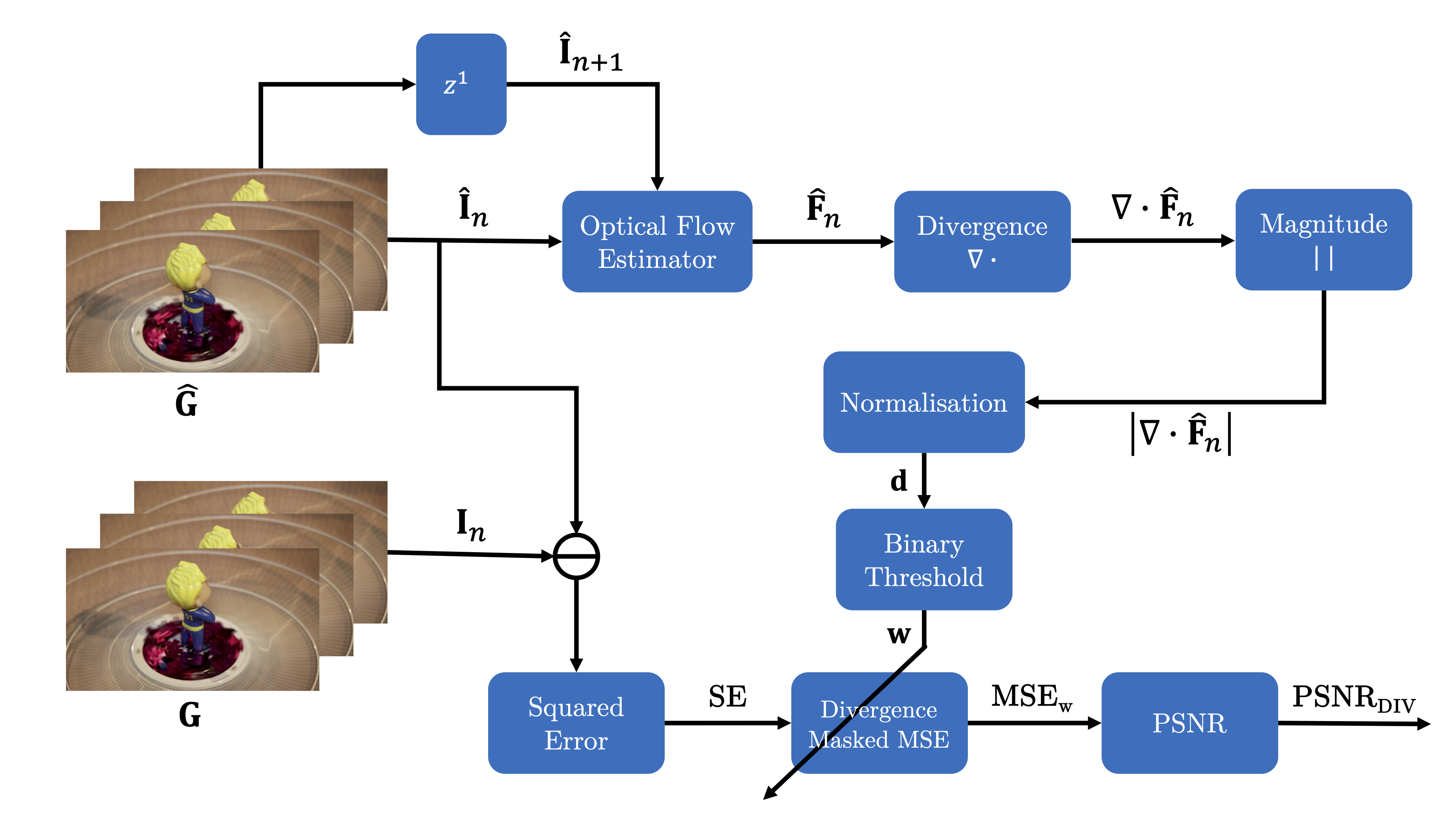}
    \caption{\textit{Flow diagram for our proposed metric. Inputs and outputs are annotated above flow arrows. Operations are denoted by blue boxes. Our metric uses the divergence of the motion field for weighting the mean squared error between pixel values of the interpolated and reference sequences.}\label{fig:flow_diagram}}
\end{figure*}

A flow diagram for our proposed VFI quality metric is shown in Figure \ref{fig:flow_diagram}. We define two video sequences. An interpolated sequence $\hat{\mathbf{G}}$ which has been interpolated from a low frame rate to a higher frame rate, and a reference sequence $\mathbf{G}$ which is the ground truth target high frame rate video. For a current time $n$, we have an interpolated frame, $\hat{\mathbf{I}}_n\in\hat{\mathbf{G}}$, and an interpolated frame $\hat{\mathbf{I}}_{n+1}\in\hat{\mathbf{G}}$ at the successive future time $n+1$. The motion field between these two frames $\hat{\mathbf{F}}_n = \left[\hat{\mathbf{u}}(\mathbf{x}), \hat{\mathbf{v}}(\mathbf{x})\right]$ is computed. The divergence of this motion field can be calculated from it's two components as follows.

\begin{equation}
    \nabla\cdot \hat{\mathbf{F}}_n(\mathbf{x}) = \frac{\partial \hat{\mathbf{u}}(\mathbf{x})}{\partial x} + \frac{\partial \hat{\mathbf{v}}(\mathbf{x})}{\partial y}
\end{equation}

In practice the absolute value of $\nabla\cdot\hat{\mathbf{F}}_n$ is taken as we simply want to detect motion field singularities. This divergence magnitude is normalised between $[0,1]$ by dividing by the maximum value for our motion field at time $n$.

\begin{equation}
    \mathbf{d}(\mathbf{x}) = \frac{\left|\nabla\cdot \hat{\mathbf{F}}_n(\mathbf{x})\right|}{\max\left(\left|\nabla\cdot \hat{\mathbf{F}}_n(\mathbf{x})\right|\right)}
\end{equation}

Using this normalised value $\mathbf{d}(\mathbf{x})$ allows us to more consistently assess importance of local motion field singularities. Essentially regions where $\mathbf{d}(\mathbf{x}) \approx 0$ can be taken to be temporally smooth. This can be used as a good proxy to indicate regions of a frame where humans may notice motion errors. We use a simple binary mask, $\mathbf{w}(\mathbf{x})$, based on a pre-defined threshold of $\mathbf{d}(\mathbf{x})$, T (we use $\text{T}=0.01$, based on our parameter optimisation in Section \ref{sec:mot-est-thresh-val}), to segment our motion field into regions of high divergence and low divergence.

\begin{equation}
\mathbf{w}(\mathbf{x}) = 
\begin{cases} 
      0 & \mathbf{d(\mathbf{x})} \leq \text{T} \\
      1 & \text{Otherwise}
\end{cases}
\end{equation}

We then take the squared error between the interpolated frame $\hat{\mathbf{I}}_{n}$ and the corresponding reference frame $\mathbf{I}_{n}\in\mathbf{G}$. This binary mask along with the squared error is used to calculate a weighted mean squared error (MSE), referred to as divergence masked MSE (MSE\textsubscript{w}). The normalising factor for the MSE is based on the sum of the number of pixels where $\mathbf{d}(\mathbf{x})$$\geq$T, denoted as $Z=\sum_\mathbf{x}\mathbf{w}(\mathbf{x})$. This results in only calculating image errors for regions of temporal inconsistency.

\begin{equation}
 \text{MSE}_\text{w} = \frac{1}{Z}\sum_\mathbf{x}{\mathbf{w}(\mathbf{x})\left( \mathbf{I}_n(\mathbf{x})-\hat{\mathbf{I}}_n(\mathbf{x})\right)^2}
\end{equation}

This modified MSE can simply be used in the conventional expression for PSNR, resulting in our VFI quality metric PSNR\textsubscript{DIV}.

\begin{equation}
 \text{PSNR}_\text{DIV} = 20 \cdot \log_{10}\left( \frac{255}{\sqrt{\text{MSE}_\text{w}}}\right)
\end{equation}

Concisely, our metric is a modified PSNR that uses MSE weighted by motion divergence, which is a measure of temporal inconsistency.

\section{Experimental Setup}
\label{setup}

In this section, we describe the dataset and statistical measures used to evaluate and compare the performance of our proposed metric.

\subsection{Evaluation Dataset} 

We evaluate the proposed method using the BVI-VFI database \cite{danier2023bvi}, the only publicly available subjective quality dataset containing uncompressed video sequences, in a 8-bit YUV420 format, with distortions exclusively induced by video frame interpolation (VFI). Compared to the two other most commonly used VFI quality datasets, VFIPS \cite{hou2022perceptual} and VFIIQA \cite{han2025full}, it is the only one that reports Differential Mean Opinion Scores (DMOS) making it the most suitable for FR metric assessment. The full BVI-VFI dataset spans three spatial resolutions (540p, 1080p, 2160p) and three frame rates (30, 60, 120fps), offering diverse content to study perceptual artefacts. For this work, we focus on a subset comprising 540p and 1080p resolutions at 30 and 60fps, as these are more common resolution and framerate conversion scenarios in addition to being more challenging for quality metrics. 

The subset includes 24 source sequences, each processed by five representative VFI methods at two different frame rates (in total this gives 240 scored sequences): basic frame averaging/repeating (to simulate motion blur and judder), motion-based deep learning approaches (DVF \cite{liu2017video} and QVI \cite{xu2019quadratic}), and the state-of-the-art kernel-based ST-MFNet \cite{danier2022stmfnet}, which combines deformable convolutions with texture synthesis based post-processing. These methods collectively span classical and modern VFI algorithms, enabling comprehensive analysis of artefact types. For each interpolated sequence BVI-VFI provides DMOS derived from a large-scale subjective study involving 189 participants, which quantify perceptual quality degradation relative to a reference sequence. DMOS is obtained by firstly computing the differential opinion score $d_{ij}$ between a subject $i$'s score $\hat{s}_{ij}$ for an interpolated video $j$ and the score for its corresponding reference video $s_{ij}$ assigned by the same participant.

\begin{equation}
d_{ij} = s_{ij} - \hat{s}_{ij}
\end{equation}

The DMOS for each interpolated video is obtained by averaging the differential opinion scores for each of the $N_j$ participants. Each video in the BVI-VFI dataset has been scored by at least 20 participants ($N_{j}\geq20$).

\begin{equation}
\text{DMOS}_j = \frac{1}{N_j}\sum^{N_j}_{i=1}{d_{ij}}
\end{equation}

BVI-VFI provides these conventional DMOS values along with ones that have been post-processed to P.910 specifications \cite{itu2000recommendation}, which accounts for biases in the scoring of participants and re-weights the differential opinion scores based on this. We use these de-biased DMOS values to evaluate our metric.

\subsection{Compared Quality Metrics} 
We benchmark our metric against widely adopted quality metrics and specialized VFI quality assessors. Notably few FR metrics have been developed since FloLPIPS. Recently there has been a resurgence in VFI quality metric development\cite{han2025full,ccokmez2025clip}. As of writing, weights are yet to be released for these models, however we compare against the same metrics evaluated in those papers which all happen to be pre-2022. PSNR and SSIM are included as foundational baselines, measuring pixel-wise fidelity and structural similarity, respectively. For PSNR and PSNR\textsubscript{DIV}, we compute scores solely on the luminance channel (Y). Gradient Magnitude Similarity Deviation (GMSD) \cite{xue2013gradient} is included, leveraging spatial gradient deviations to capture edge and texture degradation. We include LPIPS, a deep perceptual metric that compares activated network features between reference and interpolated frames. Although these image metrics lack temporal modelling, we include them to maintain consistency with prior VFI literature. We use FloLPIPS as a representative VFI video quality metric that incorporates temporal information. FloLPIPS is used with AlexNet \cite{krizhevsky2012alex} as its feature extractor (we also use LPIPS with AlexNet) and PWC-Net as its motion estimator \cite{sun2018pwc}.

As an ablation, we adapt FloLPIPS’s motion error weighting scheme (\ref{eq:flow-weight}) to MSE, creating a baseline (PSNR\textsubscript{EPE}) where motion field EPE weights pixel-wise errors. Which is normalised by the sum of the differences between the motion fields. We take this as the weighting function for MSE and refer to it as PSNR\textsubscript{EPE}.

\subsection{Evaluation Metrics}
To evaluate the alignment between objective quality metrics and subjective judgments, we employ the four statistical measures previously mentioned in Section \ref{sec:vfi-qual-metrics} which are PLCC, SRCC, KRCC, and RMSE.

We follow the VQEG recommendations for fitting quality metric values to perceptual scores \cite{vqeg2000final}. A logistic function $Y(x)$ with four-parameter ($\beta_{1-4}$) is fitted to map metric values to DMOS scores using initial parameter estimates from the VQEG report. 

\begin{equation}
    Y(x) = \beta_2 + \frac{\beta_1-\beta_2}{1+\exp\left(-\frac{x-\beta_3}{|\beta_4|}\right)}
\end{equation}

The optimization process combines exhaustive search for coarse tuning with the limited-memory Broyden-Fletcher-Goldfarb-Shanno (L-BFGS) algorithm in SciPy\footnote{\url{https://docs.scipy.org/doc/scipy/reference/optimize.minimize-lbfgsb.html}} for refinement of the logistic fit parameters.

\section{Results and Discussion}
\label{sec:eval}

\begin{table*}[]
\centering
\caption{The performance of each quality assessment model in predicting DMOS on the BVI-VFI dataset. For each statistical metric, the best and second best results are \textbf{bolded} and \underline{underlined} respectively.\label{tab:results}}

\begin{tabular}{@{}c||cccc||cccc||cccc@{}}
\toprule
                   & \multicolumn{4}{c||}{540p}                                     & \multicolumn{4}{c||}{1080p}                                    & \multicolumn{4}{c}{\multirow{2}{*}{Overall}} \\ \cmidrule(lr){2-9}
                   & \multicolumn{2}{c|}{15fps$\rightarrow$30fps}       & \multicolumn{2}{c||}{30fps$\rightarrow$60fps} & \multicolumn{2}{l|}{15fps$\rightarrow$30fps}       & \multicolumn{2}{c||}{30fps$\rightarrow$60fps} & \multicolumn{4}{c}{}                         \\ \midrule
Metric             & PLCC & \multicolumn{1}{c|}{SRCC} & PLCC         & SRCC        & PLCC & \multicolumn{1}{c|}{SRCC} & PLCC         & SRCC        & PLCC      & KRCC      & SRCC     & RMSE      \\ \midrule
PSNR               & 0.41 & \multicolumn{1}{l|}{0.27} & 0.35         & 0.45        & 0.40 & \multicolumn{1}{l|}{0.45} & 0.61         & 0.66        & 0.42      & 0.32      & 0.46     & 16.70     \\
SSIM               & 0.40 & \multicolumn{1}{l|}{0.35} & 0.48         & 0.60        & 0.42 & \multicolumn{1}{l|}{0.44} & 0.62         & 0.71        & 0.48      & 0.36      & 0.53     & 16.06     \\
GMSD               & 0.48 & \multicolumn{1}{l|}{0.41} & \underline{0.56}         & \textbf{0.67}        & 0.56 & \multicolumn{1}{l|}{\textbf{0.58}} & 0.68         & 0.73        & \underline{0.59}      & \underline{0.44}      & \textbf{0.63}     & \underline{14.78}     \\
LPIPS              & 0.38 & \multicolumn{1}{l|}{0.29} & 0.44         & 0.52        & 0.53 & \multicolumn{1}{l|}{0.45} & 0.66         & 0.66        & 0.51      & 0.36      & 0.52     & 15.83     \\
FloLPIPS           & \underline{0.49} & \multicolumn{1}{l|}{\underline{0.42}} & 0.53         & 0.55       & \underline{0.60} & \multicolumn{1}{l|}{0.55} & \underline{0.78}        & \underline{0.74}        & 0.58     & 0.40      & \underline{0.58}     & 14.98     \\ \midrule
PSNR\textsubscript{EPE} & 0.36 & \multicolumn{1}{l|}{0.31} & 0.47         & 0.50        & 0.44 & \multicolumn{1}{l|}{0.34} & 0.44         & 0.34        & 0.38      & 0.24      & 0.35     & 16.96     \\
PSNR\textsubscript{DIV}     & \textbf{0.52} & \multicolumn{1}{l|}{\textbf{0.45}} & \textbf{0.65}         & \underline{0.63}        & \textbf{0.61} & \multicolumn{1}{l|}{\underline{0.57}} & \textbf{0.80}         & \textbf{0.76}        & \textbf{0.67}      & \textbf{0.45}      & \textbf{0.63}     & \textbf{13.60}     \\ \bottomrule
\end{tabular}
\end{table*}

\begin{table}[]
\centering
\caption{Bootstrapped p-values using t-test comparing PSNR\textsubscript{DIV} and FloLPIPS performance. p$<$0.05 is deemed statistically significant. \label{tab:p-vals}}
\begin{tabular}{@{}c|ccc@{}}
\toprule
                    & \multicolumn{3}{c}{Subset (\%)} \\ \midrule
Statistical Measure & 30        & 50       & 70       \\ \midrule
PLCC                & 0.00      & 0.00     & 0.00     \\
KRCC                & 0.78      & 0.11     & 0.01     \\
SRCC                & 0.00      & 0.00     & 0.00     \\
RMSE                & 0.00      & 0.00     & 0.00     \\ \bottomrule
\end{tabular}
\end{table}

In this section we quantitatively evaluate the performance of our metric by comparing it with six commonly used image/video quality metrics. We also examine the effect of changing the motion estimator, and changing the divergence threshold parameter on metric performance. Finally, we compare the computational complexity of our metric compared to FloLPIPS.

\subsection{Evaluation}

The evaluation of PSNR\textsubscript{DIV} alongside other compared quality models on the BVI-VFI dataset are presented in Table \ref{tab:results}. For PSNR\textsubscript{DIV} and PSNR\textsubscript{EPE}, we compute motion fields using OpenCV’s implementation\footnote{\url{https://docs.opencv.org/3.4/d4/dee/tutorial\_optical\_flow.html}} of the F{\"a}rneback motion estimation algorithm \cite{farneback2003two}. It can be observed that PSNR\textsubscript{DIV} outperforms the other 6 quality metrics according to all four performance measurements overall.

Compared to FloLPIPS, PSNR\textsubscript{DIV} achieves a large gain in terms of PLCC and RMSE (+0.09 and -1.38 respectively). Small gains are also seen in terms of SRCC (+0.05). When broken down by framerate and resolution our metric outperforms FloLPIPS in all categories. GMSD occasionally outperforms PSNR\textsubscript{DIV} in SRCC (e.g., for high-motion sequences), likely due to its sensitivity to edge distortions and ability to capture structural degradation.

PSNR\textsubscript{DIV} is across the board better than PSNR\textsubscript{EPE}. This demonstrates the effectiveness of our motion divergence based weighting scheme. It is worth noting that the the poor performance of LPIPS suggests that the motion weighting step plays a larger role than the perceptual part of LPIPS showing the power of using motion as a quality feature.

To determine whether the performance difference between PSNR\textsubscript{DIV} and FloLPIPS is statistically significant, we employ a bootstrapping method. We repeatedly sample (with replacement) subsets of sequences from the BVI-VFI dataset, including their PSNR\textsubscript{DIV}, FloLPIPS, and corresponding DMOS scores. For each subset, we fit the DMOS logistic curve (as described earlier) for both metrics and compute four statistical measures (PLCC, SRCC, etc.). This is done for 2000 bootstrap iterations, generating distributions for each measure, from which we derive confidence intervals. Using a relative t-test, we calculate p-values for different sampling percentages (Table \ref{tab:p-vals}), with p$<$0.05 indicating statistical significance. Most measures show that PSNR\textsubscript{DIV} outperforms FloLPIPS significantly. However, KRCC improvements require larger sample sizes (70\%) to reach significance, suggesting that ranking consistency (KRCC) is more sensitive to dataset diversity than absolute error measures (PLCC/RMSE).

\begin{figure}[] 
    \begingroup
    \setlength{\tabcolsep}{0pt} 
    \begin{tabular}{cc}
       \includegraphics[width=0.5\linewidth]{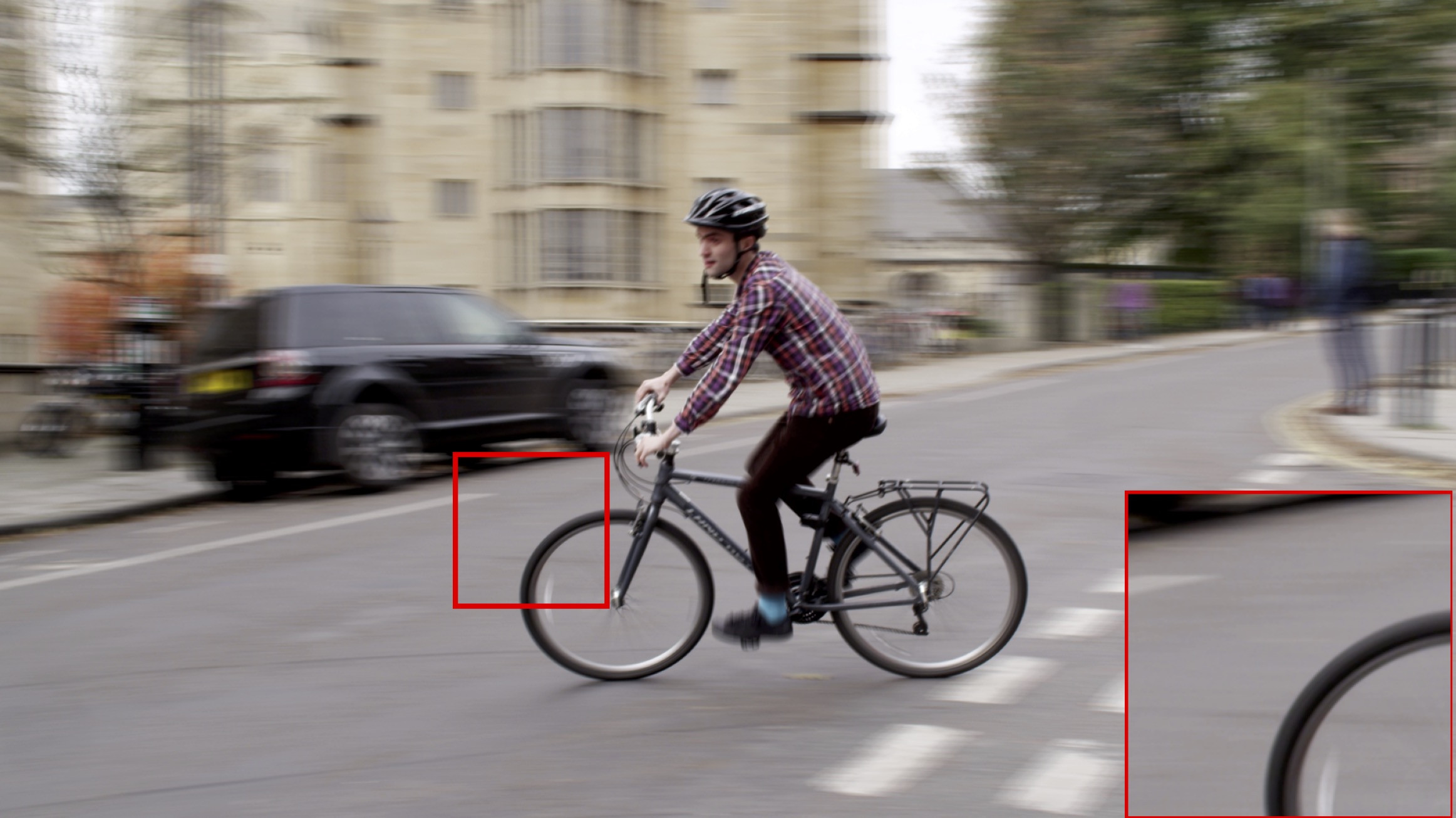}  & \includegraphics[width=0.5\linewidth]{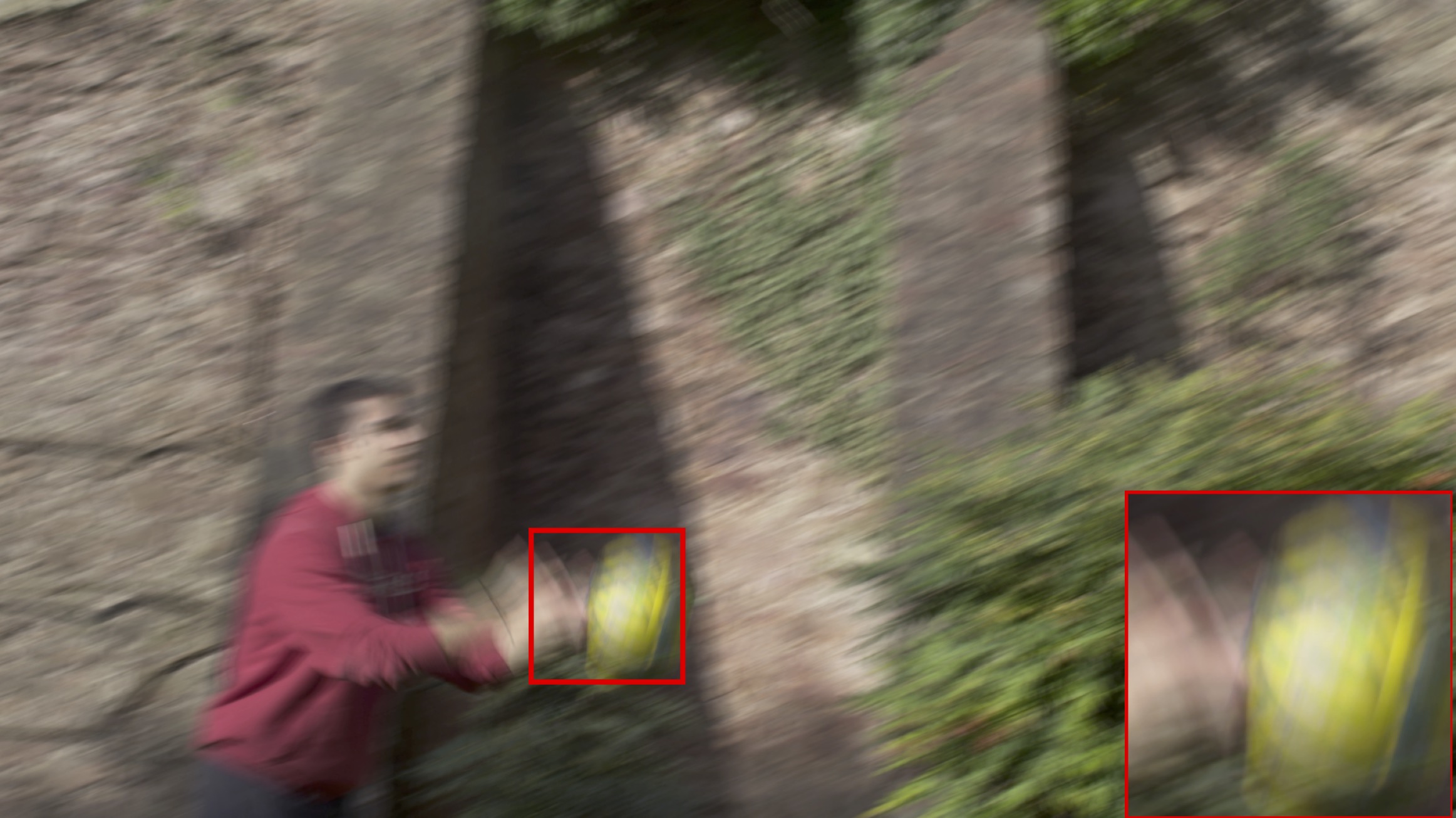} \\
       (a) & (b) \\
       \includegraphics[width=0.5\linewidth]{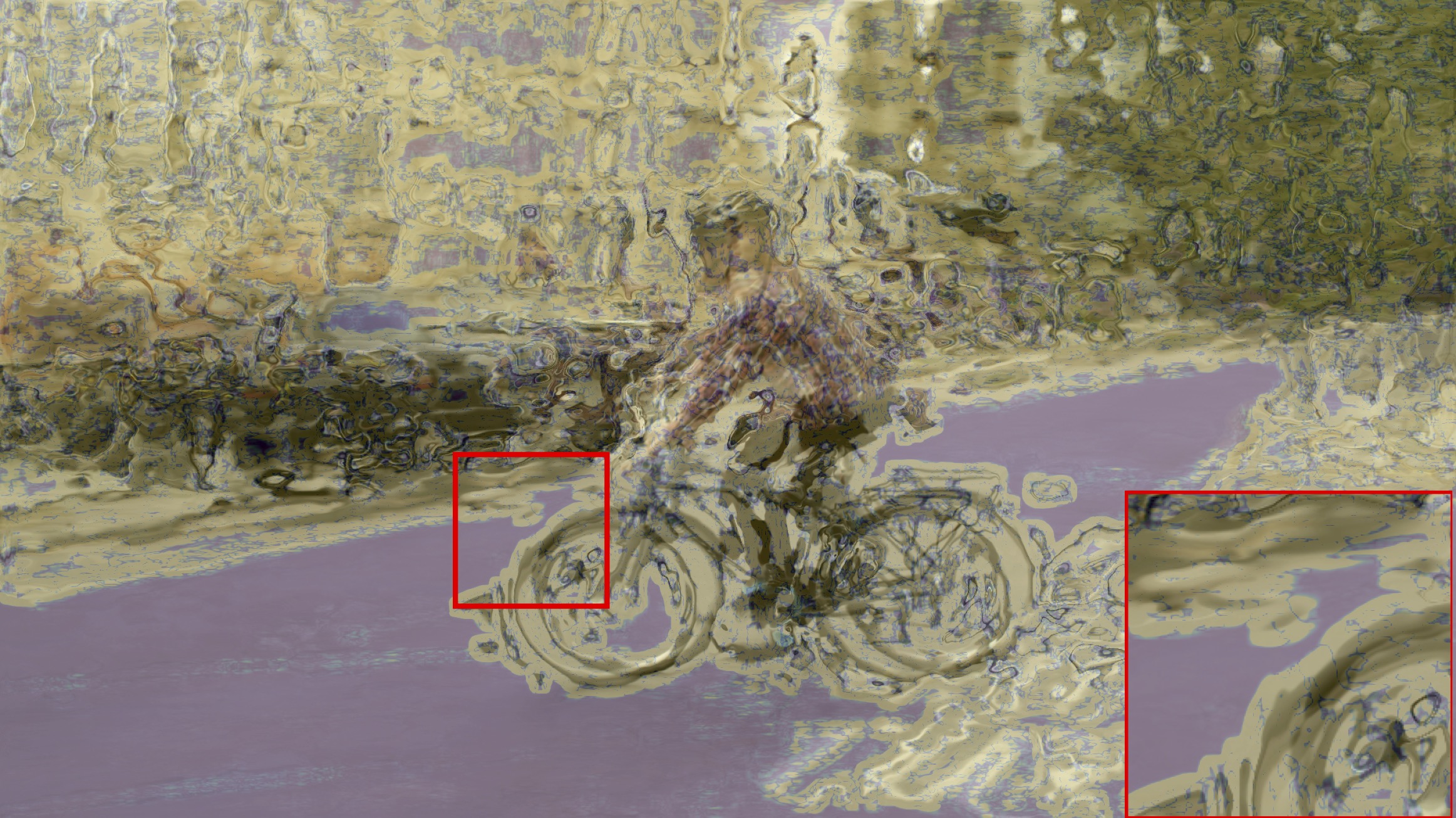}  & \includegraphics[width=0.5\linewidth]{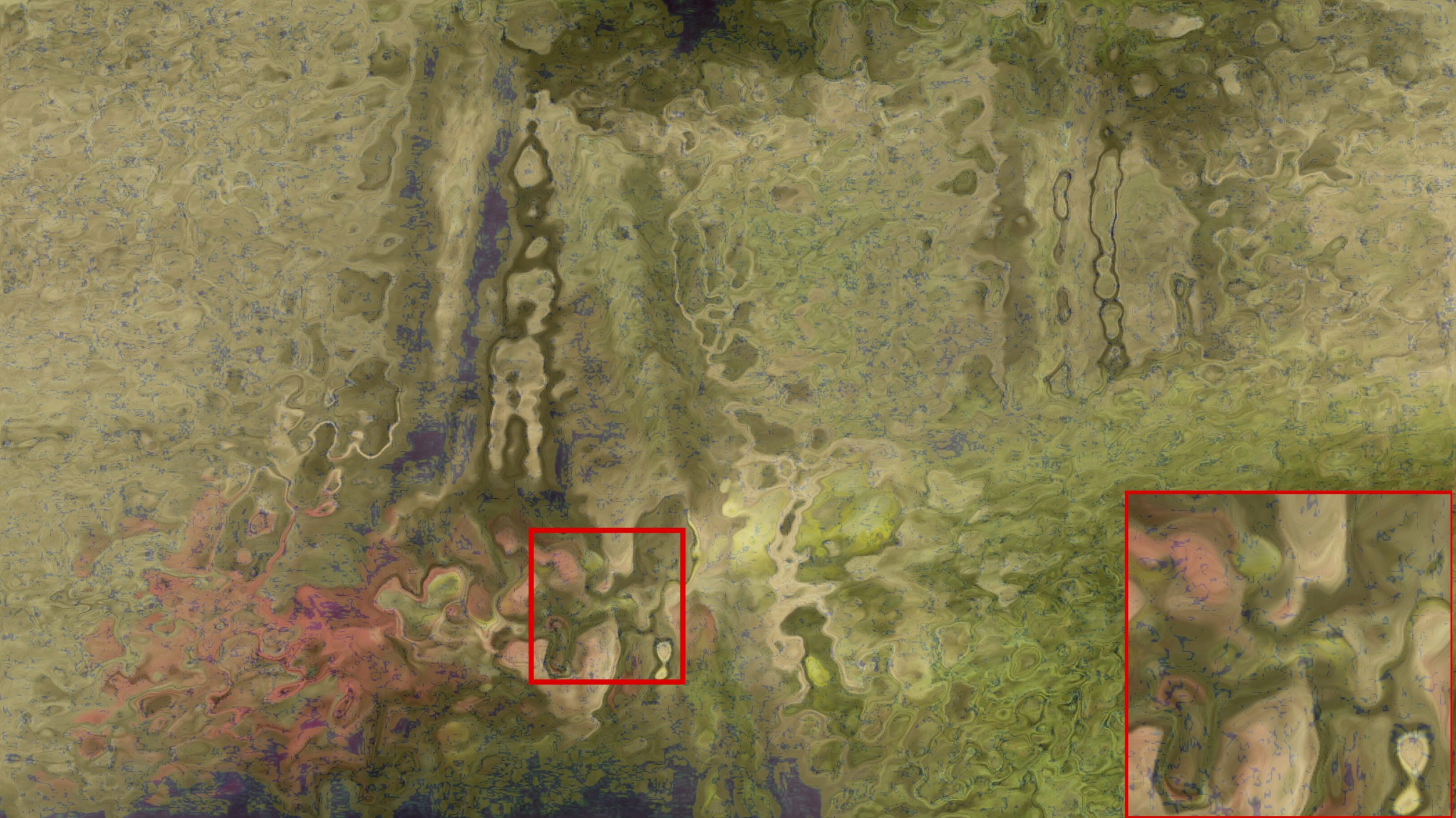} \\
       (c) & (d) \\
       \includegraphics[width=0.5\linewidth]{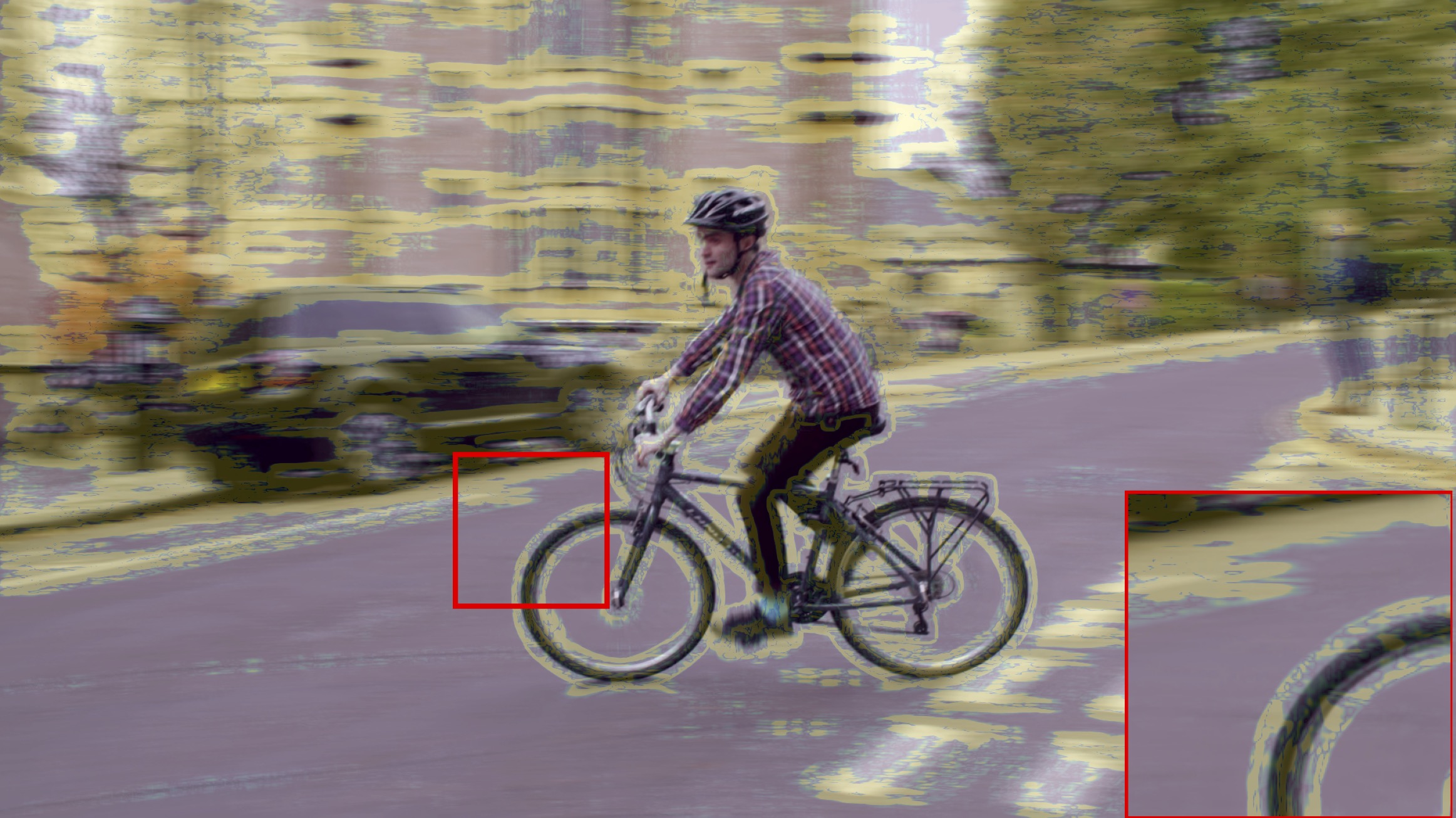}  & \includegraphics[width=0.5\linewidth]{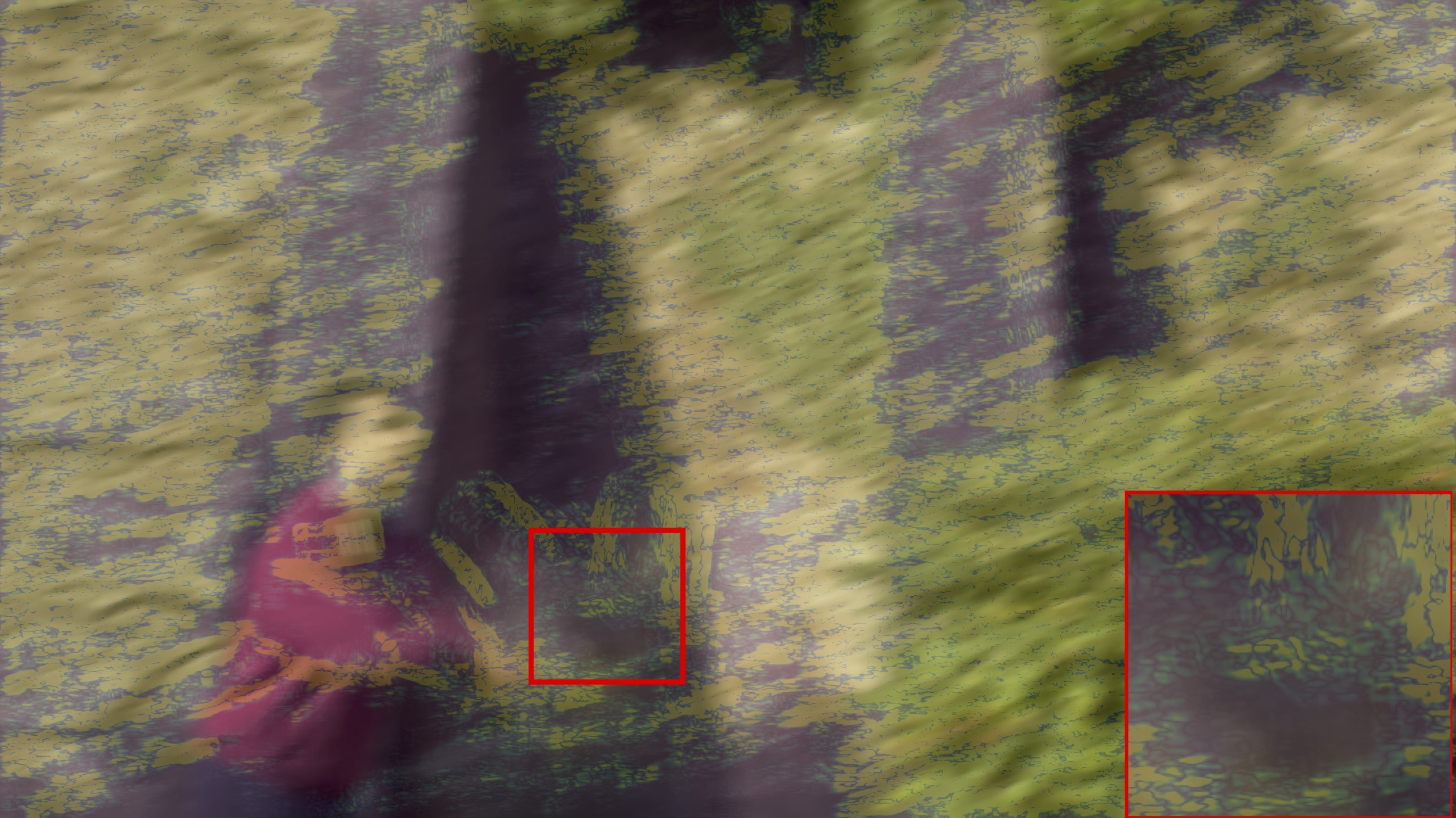} \\
       (e) & (f) \\
    \end{tabular}
    \endgroup
        
  \caption{\textit{Frames from two reference sequences in the BVI-VFI dataset are shown in (a-b). The corresponding interpolated frame generated by DVF (c-d) and ST-MFNet (e-f) are overlaid with values of our divergence measure $\mathbf{d}(\mathbf{x})$. Regions where $\mathbf{d}(\mathbf{x})\geq 0.01$ are shown in yellow and regions where $\mathbf{d}(\mathbf{x})=0$ are shown in blue. Crops are shown inset outlined in red. This demonstrates the effectiveness of divergence as a metric for detecting regions of temporal inconsistency.}}
  \label{fig:psnr-div-vis}
  
\end{figure}

A visualisation of our weighting scheme is shown in Figure \ref{fig:psnr-div-vis}. As can be seen in (c-d) our divergence based metric accurately highlights areas for which temporal inconsistency is introduced by DVF. Correspondingly fewer regions of temporal inconsistency are seen in (e-f) for the more advanced ST-MFNet interpolator, but it does still highlight regions of high motion blur in the background and occlusion (e.g. edge of bicycle wheel) where interpolation errors are most likely to occur. Notably, while ST-MFNet fails to interpolate the football as seen in (f) it does appear to create a temporally consistent interpolation compared to the warping errors introduced by DVF.

\subsection{Effect of Motion Estimator and Threshold Value}
\label{sec:mot-est-thresh-val}

\begin{table*}[]
\centering
\caption{The performance of our metric on the BVI-VFI dataset. Statistical metrics for each motion estimator are measured using the motion divergence threshold that maximises its correlation to perceptual scores. \label{tab:opt-params}}
\begin{tabular}{@{}cc||cccc||cccc||cccc@{}}
\toprule
\multirow{3}{*}{\begin{tabular}[c]{@{}c@{}}Motion \\ Estimator\end{tabular}} & \multirow{3}{*}{Threshold} & \multicolumn{4}{c||}{540p}                                     & \multicolumn{4}{c||}{1080p}                                    & \multicolumn{4}{c}{\multirow{2}{*}{Overall}} \\ \cmidrule(lr){3-10}
                                                                             &                            & \multicolumn{2}{c|}{15fps$\rightarrow$30fps}       & \multicolumn{2}{c||}{30fps$\rightarrow$60fps} & \multicolumn{2}{c|}{15fps$\rightarrow$30fps}       & \multicolumn{2}{c||}{30fps$\rightarrow$60fps} & \multicolumn{4}{c}{}                         \\ \cmidrule(l){3-14} 
                                                                             &                            & PLCC & \multicolumn{1}{c|}{SRCC} & PLCC         & SRCC        & PLCC & \multicolumn{1}{c|}{SRCC} & PLCC         & SRCC        & PLCC      & KRCC      & SRCC     & RMSE      \\ \midrule
F{\"a}rneback                                                                    & 0.01                       & 0.52 & \multicolumn{1}{c|}{0.45} & 0.65         & 0.63        & 0.61 & \multicolumn{1}{c|}{0.57} & 0.80         & 0.76        & 0.67      & 0.45      & 0.63     & 13.60     \\
PWC-Net                                                                      & 0.01                       & 0.52 & \multicolumn{1}{c|}{0.47} & 0.57         & 0.65        & 0.65 & \multicolumn{1}{c|}{0.60} & 0.83         & 0.83        & 0.67      & 0.47      & 0.65     & 13.63     \\
RAFT                                                                         & 0.001                      & 0.52 & \multicolumn{1}{c|}{0.48} & 0.56         & 0.64        & 0.70 & \multicolumn{1}{c|}{0.62} & 0.83         & 0.85        & 0.67      & 0.47      & 0.66     & 13.70     \\ \bottomrule
\end{tabular}
\end{table*}

\begin{figure}[] 
    \centering
    \begingroup
    \setlength{\tabcolsep}{0pt} 
    \begin{tabular}{c}
    \includegraphics[width=0.9\linewidth]{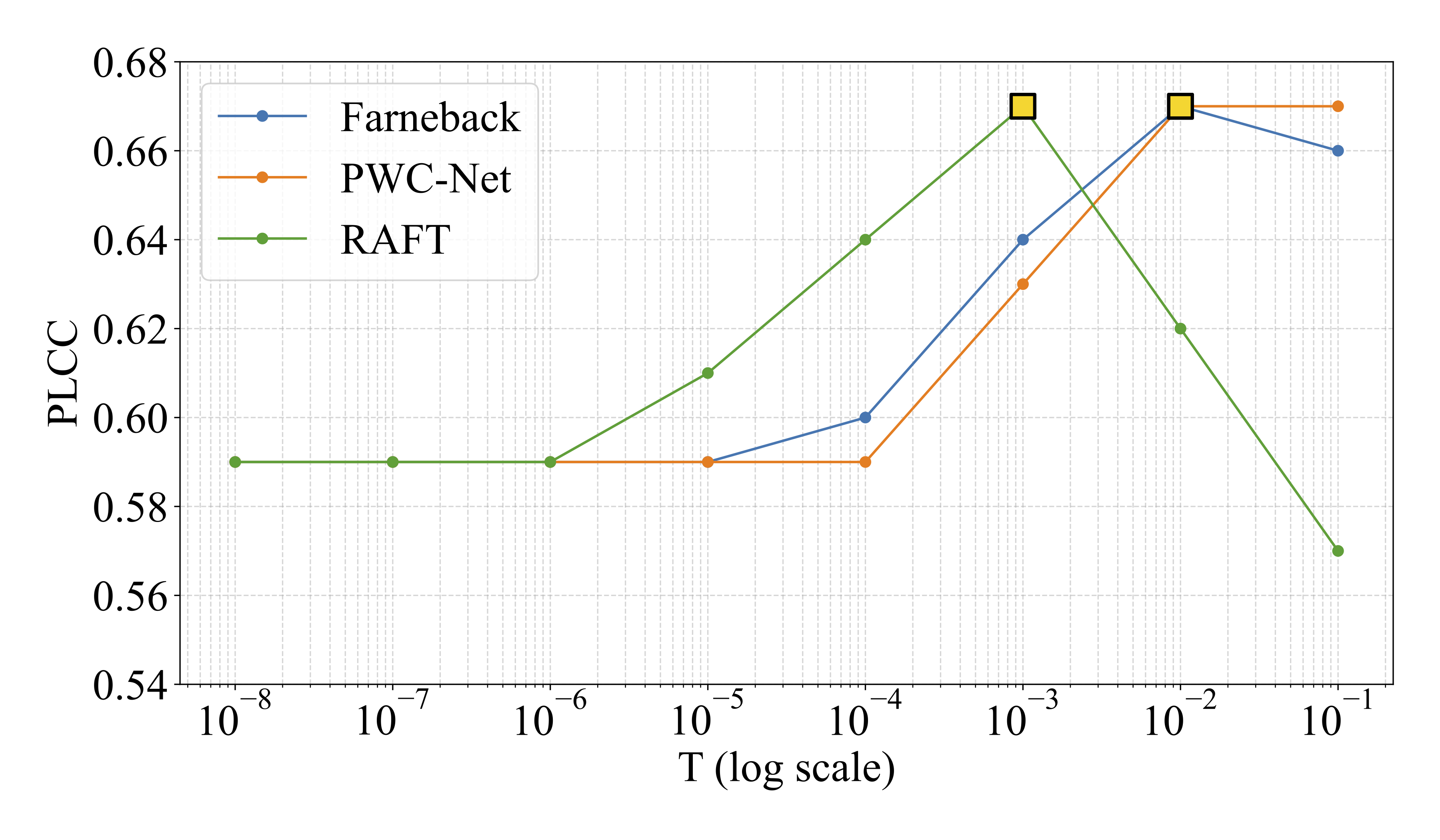} \\
    (a) \\
    \includegraphics[width=0.9\linewidth]{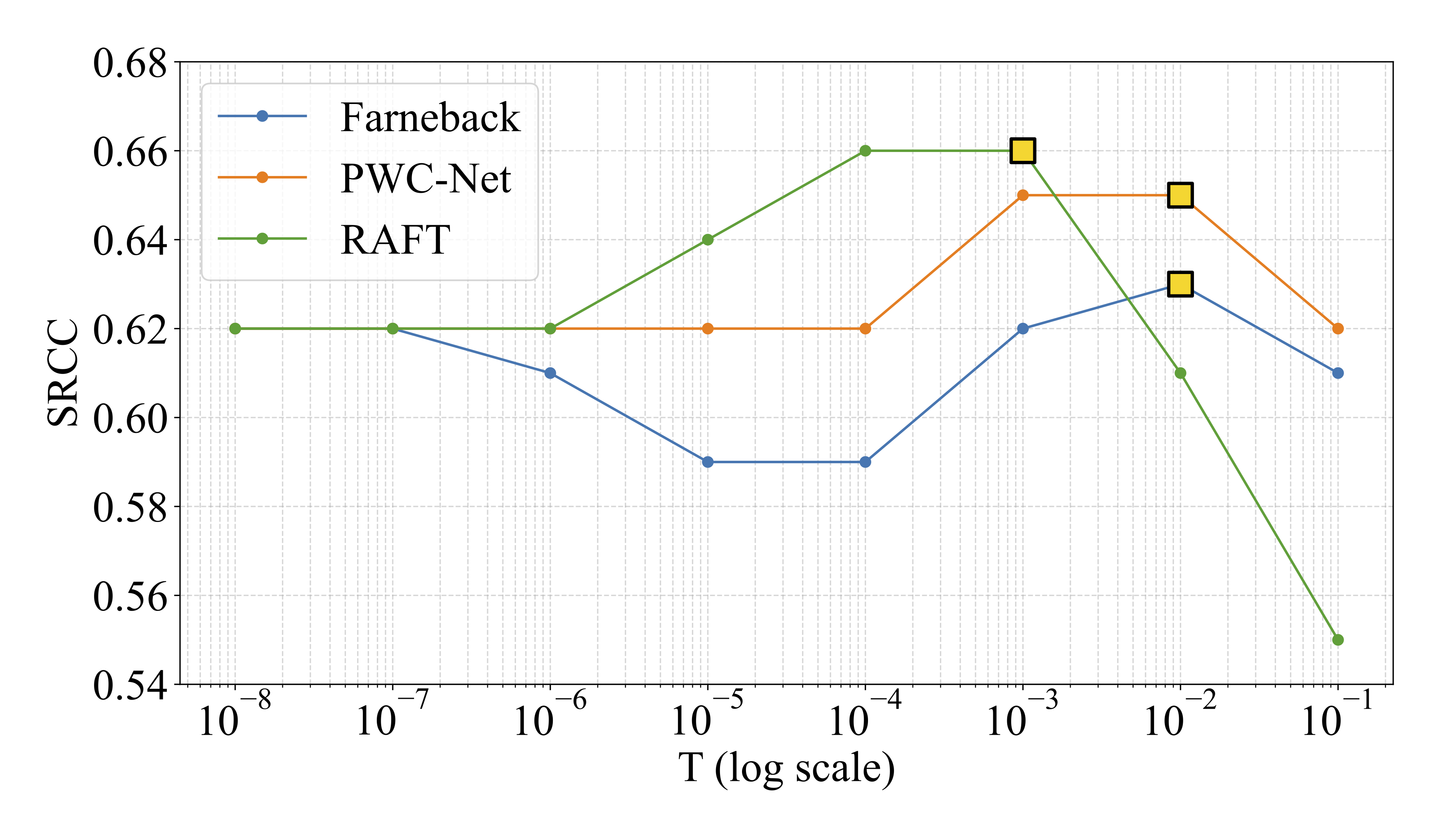} \\
    (b)
    \end{tabular}
    \endgroup
    \caption{\textit{Plots of variation in (a) PLCC and (b) SRCC with respect to motion divergence threshold (T) for different motion estimators. Values of the threshold for which correlation is maximised are marked by a yellow square.}}
    \label{fig:plcc-srcc} 
\end{figure}

To examine the effect of the choice of motion divergence threshold we performed a search across a range of values and measured the previous statistical parameters for each. In addition to make sure our metric is robust to choice of motion estimator we measure our metrics performance for three different motion estimators. We use the F{\"a}rneback algorithm, being one of the most popular classical dense motion estimators, PWC-Net as a fair comparison to FloLPIPS, and RAFT \cite{teed2020raft} as a more recent learning-based motion estimator. A sweep of the motion divergence threshold parameter (T) was performed across values spaced by an order of magnitude. The variation of the two most salient measures, PLCC and SRCC, for this parameter sweep are shown in Figure \ref{fig:plcc-srcc} (a-b) respectively. From these we can get an estimate of a good value of threshold T to maximise correlation.

Detailed results for the optimal parameters are shown in Table \ref{tab:opt-params}. It can be seen that for the F{\"a}rneback and PWC-Net motion estimators the same threshold value results in the best correlation for both of these methods with a PLCC of 0.67 and SRCC values in the 0.63-0.65 range. RAFT’s occlusion handling reduces spurious divergence in motion fields, allowing a lower threshold to capture true artefacts without over-penalizing valid interpolations. Looking at the overall results it appears that it is better to use a deep learning based motion estimator. Notably our metric when using the same motion estimator as FloLPIPS is still able to maintain good correlation overall and across individual framerate resolution subsets.

\subsection{Computational Complexity}
All metrics were evaluated on a 12th Gen Intel Core i7-12700K CPU (32GB RAM) and an NVIDIA RTX A4000 GPU. For fairness, both PSNR\textsubscript{DIV} and FloLPIPS used PWC-Net for motion estimation. We use the implementation of FloLPIPS provided by Danier \textit{et. al}\footnote{\url{https://github.com/danier97/flolpips}}. We measure memory usage using the maximum memory allocation measurement provided by PyTorch\footnote{\url{https://docs.pytorch.org/docs/stable/generated/torch.cuda.max_memory_allocated.html}}. As Table~\ref{tab:memreqs} shows, PSNR\textsubscript{DIV} reduces memory usage by 4$\times$ (3.7 GB vs. 15.2 GB) and runtime by 2.5$\times$ (124 ms vs. 305 ms per frame). With further optimization of the implementation, PSNR\textsubscript{DIV} could approach real-time evaluation (33 ms/frame), enabling integration as a loss function for VFI DNN training.

\begin{table}[h]
\centering
\caption{Memory and runtime requirements for quality evaluation of one frame from a 1080p sequence. \label{tab:memreqs}}
\begin{tabular}{@{}ccc@{}}
\toprule
Metric   & Max. Memory Usage (GB) & Runtime (ms) \\ \midrule
FloLPIPS                & 15.2          & 305        \\
PSNR\textsubscript{DIV} & 3.9           & 124        \\ \bottomrule
\end{tabular}
\end{table}

\section{Conclusions}

In this paper, we introduced PSNR\textsubscript{DIV}, a full reference video quality assessment metric tailored towards video frame interpolation quality assessment. Our metric builds on a method used to detect motion smoothness in archival footage restoration. We use this to weight MSE and calculate a modified PSNR that is based on areas with high temporal inconsistency. We evaluated our metric on the BVI-VFI subjective dataset. We show that this metric is competitive with a popular deep learning based VFI metric and can correlate better with human perception of interpolated content quality. Our metric is also 2.5$\times$ faster than FloLPIPS while using 4$\times$ less memory. Our metric is likely to achieve even greater efficiency with an optimised implementation. This makes it quite suitable for use as a loss function.

\bibliographystyle{IEEEtran}
\bibliography{refs}

\end{document}